\documentclass[12pt]{iopart}

\usepackage{iopams}

\newcommand{\prt}{\partial}

\begin{document}

\title[Evolution of transonicity in an accretion disc]
{Evolution of transonicity in an accretion disc} 

\author{Arnab K Ray$^1$ and Jayanta K Bhattacharjee$^2$}
\address{$^1$ Inter--University Centre for Astronomy and Astrophysics, 
Post Bag 4, Ganeshkhind, Pune 411007, India}
\ead{akr@iucaa.ernet.in}

\address{$^2$ Department of Theoretical Physics, 
Indian Association for the Cultivation of Science,
Jadavpur, Kolkata 700032, India}
\ead{tpjkb@mahendra.iacs.res.in}

\begin{abstract}
For inviscid, rotational accretion flows driven by a general 
pseudo-Newtonian potential on to a Schwarzschild black hole, 
the only possible fixed points are saddle points and centre-type 
points. For the specific choice of the Newtonian potential, 
the flow has only two critical points, of which the outer one 
is a saddle point while the inner one is a centre-type point.
A restrictive upper bound is imposed on the admissible range 
of values of the angular momentum of sub-Keplerian flows through 
a saddle point. These flows are very unstable to any deviation 
from a necessarily precise boundary condition. 
The difficulties against the physical 
realisability of a solution passing through the saddle point 
have been addressed through a temporal evolution of the flow,
which gives a non-perturbative mechanism for selecting 
a transonic solution passing through the saddle point. 
An equation of motion for a real-time 
perturbation about the stationary flows reveals a very close 
correspondence with the metric of an acoustic black hole, which 
is also an indication of the primacy of transonicity. 
\end{abstract}


\maketitle

\section{Introduction}
\label{sec1}

In accretion processes critical (transonic) flows --- flows which are 
regular through a critical point --- are of great interest~\cite{skc90}. 
Transonicity would imply that the bulk velocity of the
flow would be matched by the speed of acoustic propagation in the
accreting fluid. In this situation a subsonic-to-supersonic
transition or vice versa can take place in the flow either
continuously or discontinuously. In the former case the flow is
smooth and regular through a critical point for transonicity
(more specially, this can be a sonic point, at which the velocity
of the bulk flow exactly matches the speed of sound), while in the 
latter case, there will arise a shock~\cite{c89}.
The possibility of both kinds of transition
happening in an accreting system is very much real, and much
effort has been made so far in studying these phenomena.
For accretion on to a black hole especially, the argument that
the inner boundary condition at the event horizon will lead to
the exhibition of transonic properties in the flow, has been
established well~\cite{skc90}.

A paradigmatic astrophysical example of a transonic flow is the
Bondi solution in steady spherically symmetric
accretion~\cite{skc90,bon52}. What is striking about the Bondi solution
is that while the question of its realisability is susceptible to great
instabilities arising from infinitesimal deviations from an absolutely
precise prescription of the boundary condition in the steady limit of
the hydrodynamic flow, it is easy to lock on to the Bondi solution when
the temporal evolution of the flow is followed~\cite{rb02}. This dynamic 
and non-perturbative selection mechanism of the transonic solution 
agrees quite closely with Bondi's conjecture that it is the criterion 
of minimum energy that will make the transonic solution the favoured 
one~\cite{bon52,gar79}.

The appeal of the spherically symmetric flow, however, is limited by
its not accounting for the fact that in a realistic
situation, the infalling matter would be in possession of angular
momentum --- and hence the process of infall should lead to the formation
of what is known as an accretion disc, i.e. the accretion process would
be axisymmetric in nature. 
A substantial body of work over the past many years, has argued
well the case for transonicity in axisymmetric
accretion~\cite{skc90,az81,fuk87,ky94,yk95,par96,la97,lyyy97,pa97,
das02,bdw04,das04,abd06,dbd06}.
Since in either of the two kinds of astrophysical flows --- spherically 
symmetric or axisymmetric --- transonic solutions pass through a critical 
point, it should be
important to have a direct understanding of the nature of the critical
points of the flow (and, of course, the physical solutions which pass
through them). Without having to take recourse to the conventional
approach of numerically integrating the governing non-linear flow
equations, an alternative approach would be to adopt mathematical 
methods from the study of dynamical systems~\cite{js99}. In accretion 
literature one would come across some relatively recent works which 
have indeed made use of the techniques of dynamical 
systems~\cite{rb02,ap03,crd06}. Through this approach
a complete and mathematically rigorous prescription can be made
for the exclusive nature of the critical points in axisymmetric
pseudo-Schwarzschild accretion, and, in consequence, the possible
behaviour of the flow solutions passing through those points
and their immediate neighbourhood. In conjunction with a knowledge
of the boundary conditions for physically feasible inflow solutions,
this makes it possible to form an immediate qualitative notion
of the solution topologies. What is more, this study has been
shown to set a more restrictive condition on the specific angular
momentum of sub-Keplerian flows passing through the critical points.

While mentioning these general issues, it must be stressed however, 
that the specific purpose of the present work is to examine for the 
case of disc accretion some aspects of the feasibility of its transonic 
solutions, particularly their stability against the choice of a boundary 
condition and their long-time evolutionary properties on large length 
scales. To do so it would be necessary first to have a clear notion of 
at least the qualitative features of the phase portrait of the steady 
flow solutions, and its critical points. For the pedagogically simple
and particular 
case of an axisymmetric flow driven by the classical Newtonian potential, 
it has been shown here that there are only two
critical points for the steady flow, of which, for realistic
boundary conditions, the outer one is a saddle point, while the inner
one is a centre-type point. As an aside, 
it should be interesting here to note that this apparently simplistic
scenario is identically reproduced for an axisymmetric accretion flow,
under a fully rigorous general relativistic formalism, for certain
values of the flow parameters~\cite{dbd06}. That these 
qualitatively physical conclusions about the flow could be drawn
without falling back on the usual practice of a
numerical integration of the steady flow equations, amply
demonstrates the simplicity, the elegance and the power of the
dynamical systems approach that has been adopted to study the thin
accretion disc.

Having gained an understanding of the nature of the critical points 
in the phase portrait, the question that is then taken up is about the
preference of the accreting system for any particular velocity profile 
in the stationary limit of the flow, and a selection criterion thereof. 
In this regard the transonic solution has always been the favoured 
candidate. However, it is common knowledge from the study of dynamical 
systems, that a flow solution passing through a saddle point (the 
transonic flow in this case) cannot be realised physically~\cite{js99}. 
To address this issue satisfactorily it must be
appreciated that the real physical flow is not static in nature, but
has an explicit time-dependence. With respect to this point it is
tempting to subject the steady flow solutions to small perturbations
in real time, and then study their behaviour. This has been done
elsewhere~\cite{crd06,ray03} for the inviscid disc, and it has been shown
that the steady inflow solutions of abiding interest are all stable
under the influence of a linearised time-dependent perturbation on the
mass inflow rate.

Since one way or the other, no direct conclusion could be drawn about the
selection of a particular solution through a perturbative technique,
one could then try to have an understanding of a
true selection mechanism and the attendant choice of a
particular solution, by studying the evolution of the accreting
system through real time. A model analog shows that it is indeed possible
in principle for the temporal evolution to allow for the selection
of an inflow solution that passes through the saddle point, and under
restricted conditions it has been demonstrated that the selection criterion
conforms to Bondi's minimum energy argument, which is invoked to favour
the transonic solution in the spherically symmetric case. Interestingly
enough in this context, it has also been shown that although the perturbative
study has offered no direct clue about the selection of a solution, the
equation governing the propagation of an acoustic disturbance in the
flow bears a close resemblance to the effective metric of an acoustic
black hole. Use of this similarity has been made to argue that the flow 
would cross the acoustic horizon transonically.

\section{The equations of the flow and its fixed points} 
\label{sec2}

It is a standard practice to consider a thin, rotating, axisymmetric,
inviscid steady flow, with the condition of hydrostatic equilibrium
imposed along the transverse direction~\cite{mkfo84,fkr02}. The two
equations which determine the drift in the radial direction are
Euler's equation,
\begin{equation}
\label{euler}
v \frac{\mathrm{d}v}{\mathrm{d}r}
+ \frac{1}{\rho}\frac{\mathrm{d}P}{\mathrm{d}r}
+ \phi^{\prime}(r) - \frac{\lambda^2}{r^3} = 0  
\end{equation}
and the equation of continuity,
\begin{equation}
\label{con}
\frac{\mathrm{d}}{\mathrm{d}r}\left(\rho vrH \right) = 0  
\end{equation}
in which $\phi(r)$ is the generalised pseudo-Newtonian potential
driving the flow (with the prime denoting a spatial derivative),
$\lambda$ is the conserved angular momentum of the flow, $P$ is
the pressure of the flowing gas, and $H \equiv H(r)$ is the local
thickness of the disc~\cite{fkr02}, respectively.

The pressure, $P$, is prescribed by an equation of state for the flow.
As a general polytropic it is given as $P=K \rho^{\gamma}$, in which 
$K$ is a measure of the entropy in the flow and $\gamma$ is the
polytropic exponent. 
The function $H$ in (\ref{con}) will be determined according to the
way $P$ has been prescribed~\cite{fkr02}, while transonicity in the
flow will be measured by scaling the bulk velocity of the flow
with the help of the local speed of sound, given as
$c_{\mathrm{s}} = (\partial P/\partial \rho)^{1/2}$. 

With the polytropic relation thus specified for $P$, it is a
straightforward exercise to set down in terms of the speed of
sound, $c_{\mathrm{s}}$, a first integral of (\ref{euler}) as,
\begin{equation}
\label{eupol1st}
\frac{v^2}{2} + n c_{\mathrm{s}}^2 + \phi (r)
+ \frac{\lambda^2}{2 r^2} = \mathcal{E}  
\end{equation}
in which $n=(\gamma -1)^{-1}$, and the integration constant
$\mathcal{E}$ is the Bernoulli constant. Before moving on to
find the first integral of (\ref{con}) it should be important
to derive the functional form of $H$. Assumption of hydrostatic
equilibrium in the vertical direction deliver this form to be
\begin{equation}
\label{aitchpol}
H = c_{\mathrm{s}} \left(\frac{r}{\gamma \phi^{\prime}}\right)^{1/2}  
\end{equation}
with the help of which, the first integral of (\ref{con}) could
be recast as
\begin{equation}
\label{conpol1st}
c_{\mathrm{s}}^{2(2n +1)} \frac{v^2 r^3}{\phi^{\prime}}
= \frac{\gamma}{4 \pi^2} \dot{\mathcal{M}}^2  
\end{equation}
where $\dot{\mathcal{M}} = (\gamma K)^n \dot{m}$~\cite{skc90}
with $\dot{m}$, an integration constant itself, being physically the
matter flow rate.

To obtain the critical points of the flow, it should be necessary
first to differentiate both (\ref{eupol1st}) and (\ref{conpol1st}),
and then, on combining the two resulting expressions, to arrive at
\begin{equation}
\label{dvdrpol}
\left(v^2 - \beta^2 c_{\mathrm{s}}^2 \right)
\frac{\mathrm{d}}{\mathrm{d}r}(v^2) = \frac{2 v^2}{r}
\left[ \frac{\lambda^2}{r^2} - r \phi^{\prime}
+ \frac{\beta^2 c_{\mathrm{s}}^2}{2} 
\left(3 - r \frac{\phi^{\prime \prime}}{\phi^{\prime}}\right)\right]  
\end{equation}
with $\beta^2 = 2(\gamma +1)^{-1}$. The critical points of the flow
will be given by the condition that the entire right hand side of
(\ref{dvdrpol}) will vanish along with the coefficient of
${\mathrm{d}}(v^2)/{\mathrm{d}r}$ on the left hand side. Explicitly 
written down, following some rearrangement of terms, this will give 
the two critical point conditions as,
\begin{equation}
\label{critconpol}
v_{\mathrm{c}}^2 = \beta^2 c_{\mathrm{sc}}^2
= 2\left[r_{\mathrm{c}} \phi^{\prime}(r_{\mathrm{c}})
- \frac{\lambda^2}{r_{\mathrm{c}}^2} \right] \left[3 - r_{\mathrm{c}}
\frac{\phi^{\prime \prime}(r_{\mathrm{c}})}{\phi^{\prime}(r_{\mathrm{c}})}
\right]^{-1}  
\end{equation}
with the subscript ${\mathrm{c}}$ labelling critical point values.

To fix the critical point coordinates, $v_{\mathrm{c}}$ and $r_{\mathrm{c}}$,
in terms of the system constants, one would
have to make use of the conditions given by (\ref{critconpol}) along
with (\ref{eupol1st}), to obtain
\begin{equation}
\label{efixcrit}
\frac{2 \gamma}{\gamma -1}
\left[r_{\mathrm{c}} \phi^{\prime}(r_{\mathrm{c}})
- \frac{\lambda^2}{r_{\mathrm{c}}^2} \right] \left[3 - r_{\mathrm{c}}
\frac{\phi^{\prime \prime}(r_{\mathrm{c}})}{\phi^{\prime}(r_{\mathrm{c}})}
\right]^{-1} + \phi (r_{\mathrm{c}})
+ \frac{\lambda^2}{2 r_{\mathrm{c}}^2} = \mathcal{E}  
\end{equation}
from which it is easy to see that solutions of $r_{\mathrm{c}}$ may
be obtained in terms of $\lambda$ and $\mathcal{E}$ only, i.e.
$r_{\mathrm{c}}=f_1(\lambda, \mathcal{E})$.
Alternatively, $r_{\mathrm{c}}$
could be fixed in terms of $\lambda$ and $\dot{\mathcal{M}}$. By making
use of the critical point conditions in (\ref{conpol1st}) one could
write
\begin{equation}
\label{dotmfix}
\frac{4 \pi^2 \beta^2 r_{\mathrm{c}}^3}
{\gamma\phi^{\prime}(r_{\mathrm{c}})}
\left( \frac{2}{\beta^2}
\left[r_{\mathrm{c}} \phi^{\prime}(r_{\mathrm{c}})
- \frac{\lambda^2}{r_{\mathrm{c}}^2} \right] \left[3 - r_{\mathrm{c}}
\frac{\phi^{\prime \prime}(r_{\mathrm{c}})}{\phi^{\prime}(r_{\mathrm{c}})}
\right]^{-1}\right)^{2(n +1)} = {\dot{\mathcal{M}}}^2  
\end{equation}
with the obvious implication being that the dependence of $r_{\mathrm{c}}$
will be given as $r_{\mathrm{c}}= f_2(\lambda, \dot{\mathcal{M}})$.
Comparing these two alternative means of fixing $r_{\mathrm{c}}$, the
next logical step would be to say that for the fixed points, and for
the solutions passing through them, it should suffice to specify either
$\mathcal{E}$ or $\dot{\mathcal{M}}$~\cite{skc90}.

\section{Nature of the fixed points : A dynamical systems study}
\label{sec3}

The equations governing the flow in an accreting system are in
general first-order non-linear differential equations. There is
no standard prescription for a rigorous mathematical analysis of
these equations. Therefore, for any understanding of the behaviour
of the flow solutions, a numerical integration is in most cases
the only recourse. On the other hand, an alternative approach
could be made to this question, if the governing equations are
set up to form a standard first-order autonomous dynamical 
system~\cite{js99}. This is a very usual practice in general 
fluid dynamical studies~\cite{bdp93},
and short of carrying out any numerical integration, this approach
allows for gaining physical insight into the behaviour of the
flows to a surprising extent. As a first step towards this end,
for the stationary polytropic flow, as given by (\ref{dvdrpol}),
it should be necessary to parametrise this equation and set up
a coupled autonomous first-order dynamical system as~\cite{js99}
\begin{eqnarray}
\label{dynsys}
\frac{\mathrm{d}}{\mathrm{d}\tau}(v^2)&=& 2v^2 \left[
\frac{\lambda^2}{r^2} - r \phi^{\prime} 
+ \frac{\beta^2 c_{\mathrm{s}}^2}{2}
\left( 3 - r \frac{\phi^{\prime \prime}}
{\phi^{\prime}} \right) \right] \nonumber \\
\frac{\mathrm{d}r}{\mathrm{d} \tau}&=& r \left(v^2 -
\beta^2 c_{\mathrm{s}}^2 \right)  
\end{eqnarray}
in which $\tau$ is an arbitrary mathematical parameter. With respect
to accretion studies in particular, this kind of parametrisation has 
been reported before~\cite{rb02,ap03,crd06}, but the present treatment,
as far as disc accretion is concerned, will much more thoroughly 
highlight some serious questions about the feasibility of transonic
flows. 

The critical points have themselves been fixed in terms of the flow
constants. About these fixed point values, upon using a perturbation
prescription of the kind
$v^2 = v_{\mathrm{c}}^2 + \delta v^2$, $c_{\mathrm{s}}^2 =
c_{\mathrm{sc}}^2 + \delta c_{\mathrm{s}}^2$ and
$r = r_{\mathrm{c}} + \delta r$, one could derive a set of two
autonomous first-order linear differential equations in the
$\delta v^2$ --- $\delta r$ plane, with $\delta c_{\mathrm{s}}^2$ itself
having to be first expressed in terms of $\delta r$ and $\delta v^2$,
with the help of (\ref{conpol1st}) --- the continuity equation
--- as
\begin{equation}
\label{varsound}
\frac{\delta c_{\mathrm{s}}^2}{c_{\mathrm{sc}}^2} = - \frac{\gamma -1}
{\gamma + 1} \left( \frac{\delta v^2}{v_{\mathrm{c}}^2}
+ \left[3 - r_{\mathrm{c}}\frac{\phi^{\prime \prime}(r_{\mathrm{c}})}
{\phi^{\prime}(r_{\mathrm{c}})} \right] \frac{\delta r}{r_{\mathrm{c}}}
\right) . 
\end{equation}
The resulting coupled set of linear equations in $\delta r$ and
$\delta v^2$ will be given as
\begin{eqnarray}
\label{lindynsys}
\frac{1}{{2v_{\mathrm{c}}^2}}\frac{\mathrm{d}}{\mathrm{d}\tau}
(\delta v^2) &=&  \frac{\mathcal A}{2}\left(\frac{\gamma -1}{\gamma + 1}
\right) \delta v^2 
- \Bigg [\frac{2 \lambda^2}{r_{\mathrm{c}}^3} +
\phi^{\prime}(r_{\mathrm{c}}) + r_{\mathrm{c}}\phi^{\prime \prime}
(r_{\mathrm{c}}) \nonumber \\
&+& \frac{\beta^2 {c_{\mathrm{sc}}^2}}{2}
\frac{\phi^{\prime \prime}(r_{\mathrm{c}})}{\phi^{\prime}(r_{\mathrm{c}})}
\mathcal{B} + \frac{\beta^2 {c_{\mathrm{sc}}^2}}{2}
\left(\frac{\gamma -1}{\gamma + 1} \right)
\frac{{\mathcal A}^2}{r_{\mathrm{c}}} \Bigg] \delta r \nonumber \\
\frac{1}{r_{\mathrm{c}}}\frac{\mathrm{d}}{\mathrm{d}\tau}(\delta r)
&=& \frac{2\gamma}{\gamma + 1} \delta v^2 - \mathcal{A} \left(
\frac{\gamma -1}{\gamma + 1} \right)
\frac{{v_{\mathrm{c}}^2}}{r_{\mathrm{c}}} \delta r  
\end{eqnarray}
in which
\begin{displaymath}
\label{coeffs}
\mathcal{A} = r_{\mathrm{c}}\frac{\phi^{\prime \prime}(r_{\mathrm{c}})}
{\phi^{\prime}(r_{\mathrm{c}})} - 3 \,,  \qquad
\mathcal{B} = 1 + r_{\mathrm{c}}
\frac{\phi^{\prime \prime \prime}(r_{\mathrm{c}})}
{\phi^{\prime \prime}(r_{\mathrm{c}})}
- r_{\mathrm{c}}\frac{\phi^{\prime \prime}(r_{\mathrm{c}})}
{\phi^{\prime}(r_{\mathrm{c}})} . 
\end{displaymath}
Trying solutions of the kind $\delta v^2 \sim \exp(\Omega \tau)$
and $\delta r \sim \exp(\Omega \tau)$ in (\ref{lindynsys}), will
deliver the eigenvalues $\Omega$ --- growth rates of $\delta v^2$ and
$\delta r$ --- as
\begin{eqnarray}
\label{eigen}
\Omega^2 &=& \frac{2 r_{\mathrm{c}}
\phi^{\prime}(r_{\mathrm{c}}) \beta^2 c_{\mathrm{sc}}^2}{(\gamma + 1)}
\Bigg( \left[ \left(\gamma - 1 \right)
{\mathcal A} - 2 \gamma \left(4 + {\mathcal A} \right) + 2 \gamma
{\mathcal{B}}\left(1 + \frac{3}{\mathcal A}\right) \right] \nonumber \\
&-& \frac{\lambda^2}{\lambda_{\mathrm K}^2(r_{\mathrm{c}})}
\left[4 \gamma + \left(\gamma - 1 \right){\mathcal A} + 2 \gamma
{\mathcal{B}}\left(1 + \frac{3}{\mathcal A}\right) \right] \Bigg)
\end{eqnarray}
where $\lambda_{\mathrm K}(r)$ is the local Keplerian angular momentum,
expressed as $\lambda_{\mathrm K}^2(r) = r^3 \phi^{\prime}(r)$.

Once the position of a critical point, $r_{\mathrm{c}}$, has been
ascertained, it is then a straightforward task to find the nature
of that critical point by using $r_{\mathrm{c}}$ in (\ref{eigen}).
Since it has been discussed in Section~\ref{sec2} that $r_{\mathrm{c}}$
is a function 
of $\lambda$ and $\mathcal E$ (or $\dot{\mathcal M}$) for polytropic
flows, it effectively implies that $\Omega^2$ can, in principle, be
rendered as a function of the flow parameters. 
A generic conclusion that can be drawn about the critical points from
the form of $\Omega^2$ in (\ref{eigen}), is that for a conserved
pseudo-Schwarzschild axisymmetric flow driven by any potential, the only
admissible critical points will be saddle points and centre-type points.
For a saddle point, $\Omega^2 > 0$, while for a centre-type point,
$\Omega^2 < 0$. Once the behaviour of all the physically relevant
critical points has been understood in this way, a complete qualitative
picture of the flow solutions passing through these points (if they
are saddle points), or in the neighbourhood of these points (if they
are centre-type points), can be constructed, along with an impression
of the direction that these solutions can have in the phase portrait
of the flow~\cite{js99}.

A further interesting point that can be appreciated from the derived
form of $\Omega^2$, is related to the admissible range of values for
a sub-Keplerian flow passing through a saddle point. It is self-evident
that for this kind of flow,
$(\lambda/\lambda_{\mathrm K})^2 <1$~\cite{az81}.
However, a look at (\ref{eigen}) will reveal that a more restrictive
upper bound on $\lambda/\lambda_{\mathrm K}$ can be imposed under the
requirement that $\Omega^2 > 0$ for a saddle point, and this restriction
will naturally be applicable to solutions which pass through such a
point. This is entirely
a physical conclusion, and yet its establishing has been achieved
through a mathematical parametrisation of a dynamical system.

To appreciate the practical 
usefulness of the method developed so far, it should
be worthwhile to use the simple example of the Newtonian potential, 
$\phi = - GM/r$ as a special case. From (\ref{efixcrit}) it will 
then be possible to find the spatial coordinates of the fixed 
points, which will be given by the two roots 
of a quadratic equation, whose solution will be given by 
\begin{equation}
\label{newtroots}
r_{\mathrm{c}} = \frac{(5 - 3 \gamma) GM}{10 (\gamma -1)\mathcal{E}}
\left[ 1 \pm \sqrt{1 - \frac{10 (5 - \gamma)(\gamma - 1) \lambda^2 
{\mathcal{E}}}{(5 - 3 \gamma)^2 (GM)^2}} \right]  
\end{equation}
from which a conclusion that can be drawn is that for critical 
conditions to be obtained, there will have to be real solutions for 
$r_{\mathrm{c}}$, and this could only be achieved if the condition 
\begin{equation}
\label{lamlim}
\lambda < \frac{(5 - 3 \gamma) GM}{\sqrt{10 (5 - \gamma)(\gamma -1)
\mathcal{E}}}
\end{equation}
were to be maintained. 

The properties of these two critical points can be further analysed  
with the help of the eigenvalues of the stability matrix associated 
with the critical points. Upon using the 
Newtonian potential, $\phi = - GM/r$, in (\ref{eigen}), it would
be easy to arrive at  
\begin{equation}
\label{newteigen}
\Omega^2= 2 \beta^2 c_{\mathrm{sc}}^2
\left(\frac{5- \gamma}{\gamma +1} \right)
\frac{GM}{r_{\mathrm c}} \left[\zeta 
- \frac{\lambda^2}{\lambda^2_\mathrm{K}(r_{\mathrm c})}\right]
\end{equation}
where $\zeta = (5-3 \gamma)(5- \gamma)^{-1}$ and 
$\lambda^2_\mathrm{K}(r_{\mathrm c}) = GMr_\mathrm{c}$. 

The term in the square brackets in (\ref{newteigen}) will determine 
the sign of $\Omega^2$. Knowing that two adjacent fixed points cannot
be of the same nature~\cite{js99}, i.e. they will differ in their
respective signs of $\Omega^2$, it is now evident that for the signs 
of all other factors in (\ref{newteigen}) remaining always unchanged, 
if $r_\mathrm{c}$ is the length coordinate of the outer fixed point, 
for which $(\lambda/\lambda_\mathrm{K})^2 < \zeta$, then $\Omega^2$ 
will be positive and the fixed point will be a saddle point, while 
if $r_\mathrm{c}$ gives the inner fixed point, for which 
$(\lambda/\lambda_\mathrm{K})^2 > \zeta$, then $\Omega^2$ will 
be negative and the fixed point will be centre-type. 
It is also remarkable that this apparently simplistic scenario of two
fixed points, with the outer one being a saddle point and the inner one
being centre-type, can be identically reproduced in the case of a proper
general relativistic flow on to a Schwarzschild black hole, under
certain ranges of the flow parameters~\cite{dbd06}. All the flows of 
physical interest in these cases are governed by the boundary condition 
that at large distances the drift velocity becomes vanishingly small,
while the speed of sound approaches a constant value. This knowledge 
of the boundary condition, in conjunction with the nature of the fixed
points, makes it possible to understand what the phase trajectories 
would look like.  
In a manner of speaking, the outer fixed point, 
which is a saddle point, may be dubbed the ``sonic point" because one of
the two trajectories passing through it is a transonic inflow solution,
rising from subsonic values far away from the saddle point to attain 
supersonic values on length scales which are less than that of the 
saddle point. Transonicity is to be attained when the drift velocity 
equals the speed of the propagation of an acoustic disturbance,
which, for this system is given
by $\sqrt{2}(\gamma +1)^{-1/2}c_\mathrm{s}$~\cite{ray03}. 
Dwelling on this last point in somewhat greater detail, it may be noted
that in some earlier works~\cite{c89,mkfo84} the speed of acoustic
propagation has been scaled as an ``effective" speed of sound. This
scaling has been done by a constant factor that arises due to the chosen 
geometry of the thin disc flow (especially for a disc in vertical
hydrostatic equilibrium, with the local disc height being a function
of the acoustic velocity). This leads to the Mach number of the flow 
being scaled accordingly, and the critical condition for the flow 
will consequently be achieved when this scaled Mach number becomes
unity. In the present treatment, on the other hand, the conventional
definition of the Mach number has been adhered to, and as a result 
criticality will occur when the Mach number assumes the value 
$\sqrt{2}(\gamma +1)^{-1/2}$. Either approach is entirely equivalent
to the other, if one is mindful of the fact that the {\em exact} sonic
condition (where the bulk flow velocity matches the local {\em unscaled}
speed of sound) and the critical condition differ by a constant scaling
factor only. This distinction disappears for the case of a spherically
symmetric flow and for an axisymmetric flow with a constant disc height. 
On the other hand, the situation becomes radically more complicated for
the case of a fully general relativistic flow. Here the critical Mach
number is not a global constant, but is a local function of the critical
point coordinates~\cite{dbd06}, all of which, of course, makes it very
much difficult a mathematical exercise to fix the critical point coordinates,
calculate the slope of solutions passing through the critical points
(in terms of physical relevance these points should be saddle points), 
and then numerically integrate these solutions under appropriate 
boundary conditions. 

It has been mentioned earlier that the particular properties of a 
saddle point in the phase portrait of thin disc flow solutions
will also have a bearing on the possible range of 
values that the constant specific angular momentum, $\lambda$, may be 
allowed to have. For solutions passing through the saddle point, it 
can be easily recognised from (\ref{critconpol}) that the flow will 
be sub-Keplerian~\cite{skc90,az81}. In that case the condition, 
$(\lambda/\lambda_{\mathrm{K}})^2<1$, shall hold good. In addition 
to this, for the specific case represented by (\ref{newteigen}), the 
saddle-type behaviour of the outer fixed point would also imply that 
there would be
another upper bound on $\lambda$, given by 
$(\lambda/\lambda_{\mathrm{K}})^2< \zeta$.
For the admissible range of the polytropic index $\gamma$, the possible
range of $\zeta$ would be $0< \zeta < 1/2$. This would then imply that
the latter bound on $\lambda$ would be more restrictive, as compared to 
the former. 

Saddle points are, however, inherently unstable, and to make a solution 
pass through such a point, after starting from an outer boundary 
condition, will entail an infinitely precise fine-tuning of that 
boundary condition~\cite{rb02}. This can be demonstrated through simple 
arguments. Going back to (\ref{lindynsys}), the coupled set of linear
differential equations in $\delta v^2$ and $\delta r$ can be set down as 
\begin{equation}
\label{ratio}
\frac{{\mathrm d} \left(\delta v^2 \right)}{{\mathrm d} 
\left(\delta r \right)} = \frac{{\mathrm d} 
\left(\delta v^2 \right)/{\mathrm d} \tau}
{{\mathrm d} \left(\delta r \right)/{\mathrm d} \tau} = 
\frac{{\mathcal Q}_1 \delta v^2 + {\mathcal Q}_2 \delta r}
{{\mathcal Q}_3 \delta v^2 + {\mathcal Q}_4 \delta r}
\end{equation}
in which the constant coefficients ${\mathcal Q}_1$, ${\mathcal Q}_2$,
${\mathcal Q}_3$ and ${\mathcal Q}_4$ are to be determined simply by
inspection of (\ref{lindynsys}). It is also to be easily seen that 
${\mathcal Q}_1 = - {\mathcal Q}_4$. This makes the integration of
(\ref{ratio}) a staightforward exercise that yields 
\begin{equation}
\label{hyper}
{\mathcal Q}_2
\left(\delta r \right)^2 + 2 {\mathcal Q}_1 \left(\delta v^2 \right)
\left(\delta r \right) - {\mathcal Q}_3 \left(\delta v^2 \right)^2 
+ {\mathcal C} = 0 
\end{equation}
with ${\mathcal C}$ being an integration constant. Generally speaking 
(\ref{hyper}) is the equation of a conic section in the 
$\delta v^2$ --- $\delta r$ plane. If the origin of this plane were
to be considered to have been shifted to the saddle point, then the 
condition for solutions passing through the origin,  
i.e. $\delta v^2 = \delta r = 0$, would be 
${\mathcal C} = 0$, which reduces (\ref{hyper}) to a pair of straight
lines intersecting each other through the origin itself. All other 
solutions in the vicinity of the origin will, therefore, be hyperbolic 
in nature, a fact that is given by the condition 
$({\mathcal Q}_1^2 + {\mathcal Q}_2 {\mathcal Q}_3) > 0$. 
For the case of $\phi = -GM/r$, this contention can be verified 
completely analytically, and this shows that even a very minute 
deviation from a precise boundary condition for transonicity 
(i.e. a boundary condition
that will generate solutions to pass {\em only} through the origin, 
$\delta v^2 = \delta r = 0$) will take the stationary solution far
away from a transonic state. This extreme sensitivity of transonic 
solutions to boundary conditions is entirely in keeping with the 
nature of saddle points. It may be imagined that in a proper 
astrophysical system such precise fulfillment of a boundary 
condition will make the transonic solution well-nigh physically 
non-realisable. Indeed, this difficulty, for any kind of accreting 
system, is readily appreciated by anyone trying to carry out a numerical
integration of (\ref{euler}) to generate the transonic solutions, which 
can only be obtained when the numerics is first biased in favour of
transonicity by using the saddle point condition itself as the 
boundary condition for numerical integration. 

There is another aspect to the physical non-realisability of 
transonic solutions, although this is somewhat mathematical in 
nature. Using the condition ${\mathcal C} = 0$ will make it easy
to express $\delta v^2$ in terms of $\delta r$ and vice versa. Going
back to the set of linear equations given by (\ref{lindynsys}) and 
choosing the second one of the two equations (the choice of the first
would also have led to the same result), one gets
\begin{equation}
\label{aar}
\frac{{\mathrm d} \left(\delta r \right)}{{\mathrm d} \tau}= \pm 
\sqrt{{\mathcal Q}_1^2 + {\mathcal Q}_2 {\mathcal Q}_3} \,\,\, \delta r
\end{equation}
which can be integrated for both the roots from an arbitrary
initial value of $\delta r =[\delta r]_{\rm{in}}$ lying anywhere 
on the transonic solution, to a point
$\delta r = \epsilon$, with $\epsilon$ being very close to the
critical point given by $\delta r =0$. With this it can be shown that 
\begin{equation}
\label{tau}
\tau = \pm \frac{1}{\sqrt{{\mathcal Q}_1^2 
+ {\mathcal Q}_2 {\mathcal Q}_3}}
\int_{[\delta r]_{\rm{in}}}^{\epsilon} \frac{{\mathrm d} 
\left(\delta r \right)}
{\delta r} = \pm \frac{1}{\sqrt{{\mathcal Q}_1^2 
+ {\mathcal Q}_2 {\mathcal Q}_3 }}
\ln \Bigg{\vert} \frac{\epsilon}{[\delta r]_{\rm{in}}} \Bigg{\vert}
\end{equation}
from which it is easy to see that 
$\vert \tau \vert \longrightarrow \infty$
for $\epsilon \longrightarrow 0$. This implies
that the critical point may be reached along either of the
separatrices, only after $\vert \tau \vert$ has become
infinitely large. This divergence of the parameter $\tau$ 
indicates that in the stationary regime, solutions passing
through the saddle point are not actual solutions, but
separatrices of various classes of solutions~\cite{js99}. This fact,
coupled with the sensitivity of the stationary transonic solutions
to the choice of an outer boundary condition, makes
their feasibility a seriously questionable matter. 

\section{Dynamic evolution as a selection mechanism for transonicity}
\label{sec4}

In Section~\ref{sec3} it has been demonstrated that generating a 
stationary solution through a saddle point (a transonic solution) 
will be impossible, physically speaking. Nevertheless, in accretion 
studies transonicity is not a matter of doubt. The key to this
paradox lies in considering explicit time-dependence in the flow.
This happens because the two stationary equations (\ref{euler}) 
and (\ref{con}), as it 
is very much evident from their form, are invariant under the 
transformation $v \longrightarrow - v$ (i.e. inflows and outflows
are twin solutions of the same set of stationary equations), and it 
is this invariance that gives birth to the saddle point, which is 
an intersection point of the transonic inflow and outflow solutions. 
This invariance breaks down as soon as time-dependent terms are 
introduced in the governing equations, something that is very 
easy to see from 
\begin{equation}
\label{dyneuler}
\frac{\prt v}{\prt t} + v \frac{\prt v}{\prt r}
+ \frac{1}{\rho} \frac{\prt P}{\prt r} + \phi^{\prime}(r)
-\frac{\lambda^2}{r^3}=0 . 
\end{equation}
This will obviously imply that a choice of inflows $(v <0)$ or 
outflows $(v >0)$ has to be made right at the very beginning 
(at $t=0$), and solutions generated thereafter will be 
free of all adverse conditions that one may associate with the 
presence of a saddle point in the stationary flow. 

In trying to make a time-dependent study, one could also assure 
oneself that a linearised perturbative analysis in real time cannot
give any conclusive insight about the accreting system showing any
preference for any particular solution. For inviscid, axisymmetric
accretion, this fact has been clearly established~\cite{crd06,ray03}. 
One may then say that any selection mechanism based on explicit 
time-dependence has to be non-perturbative and evolutionary in 
character. 

However, even for the simple inviscid disc system, with explicit 
time-dependence taken into consideration, the equations of a
compressible flow cannot be integrated analytically. Hence, to 
have any appreciation
of the time-evolutionary selection of the critical solution, it should
be instructive to consider an analogous model situation first. The model
system being introduced here, describes the dynamics of a field 
$y(x,t)$ that is analogous to a steadily accreting system with two
fixed points, as has been discussed so far. The details of the 
dynamic selection of the critical solutions in the model problem have
been presented in the Appendix. 

To the extent that this model has been a good representative of the
true physical situation, the whole argument for a time-dependent and 
non-perturbative method of selection developed in the Appendix, may 
now be extended to the actual problem of thin disc accretion. For a 
steady accretion disc many previous works have upheld the case for
transonicity, although without explicitly addressing the issue of what 
special physical criterion may select the transonic solution to the 
exclusion of all other possible solutions. For the case of disc
accretion on to black holes, Liang and Thompson~\cite{lt80} make a
clear point by saying that ``the solution for the radial drift velocity of
thin disk accretion onto black holes must be transonic, and is analogous to
the critical solution in spherical Bondi accretion, except for the presence
of angular momentum." For the inviscid and thin disc at least, this is 
indeed a most crucial analogy, which will make it possible to invoke 
all the physical arguments used to uphold transonicity in spherically
symmetric flows. 

Transonicity is a settled fact~\cite{bon52,gar79} in spherically
symmetric accretion, and this has been so because at every spatial point 
in the flow, velocity evolves at a much greater rate through time, in 
comparison with density, and this, therefore, excludes
the possibility of matter accumulating in regions close to the surface of
the accretor. As a result gravity wins over pressure at small distances
and the system is naturally driven towards selecting the transonic
solution. This sort of a situation will be more so obtained for the 
case of matter accreting on to
a black hole, and it then raises the question of whether or not a likewise
time-evolutionary mechanism should be at work for the selection of the
critical inflow solution in an inviscid and thin accretion disc.
This connection need not be so obvious, considering the fact that thin
disc accretion involves, apart from a very different flow geometry, a
whole range of different physical phenomena, not invoked for spherically
symmetric flows. 

To have any analytical appreciation of how the temporal evolution acts 
as a selection mechanism in disc accretion, it should be instructive to 
prescribe the Newtonian potential for $\phi$ in (\ref{dyneuler}), as 
well as recast the pressure term in it by the polytropic relation,
$P = K \rho^{\gamma}$. All of this will lead to 
\begin{equation}
\label{timdep}
{\frac{\partial v}{\partial t}}+v{\frac{\partial v}{\partial r}}
+K{\gamma}{\rho}^{\gamma -2}{\frac{\partial \rho}{\partial r}}+
{\frac{GM}{r^2}} -{\frac{\lambda^2}{r^3}} = 0
\end{equation}
for which it should be noted that the question of generating 
transonicity in the inflow solution is not going to be affected
overmuch by the specific choice of the simple Newtonian potential 
for $\phi$, especially so in the vicinity of the 
far-off outer critical (saddle) point. It would be easy to see that to 
have a solution pass through this critical point, the temporal evolution 
should proceed in such a manner, that the inflow velocity would increase 
much faster in time than density at distances on the scale of the saddle 
point, where, for sub-Keplerian
flows, gravity, going as $r^{-2}$, dominates the rotational effects,
which go as $r^{-3}$. 

But what should be the key physical criterion guiding this dynamic
selection of the transonic solution? It is to be stressed once again 
that the
selection principle would very likely be the same as in the spherically
symmetric case, where the transonic solution is chosen by dint of its
corresponding to a configuration of the lowest possible
energy~\cite{gar79}. To have an understanding of this, an analytical
treatment of (\ref{timdep}) may be carried out, with the density gradient 
being neglected as a working approximation. On large length scales, 
where the flow is highly subsonic, this is an especially effective 
approximation, and it will allow for treating the evolution of the 
velocity field to be largely independent of the density evolution. 
The resulting non-linear partial differential equation for velocity 
is then to be integrated by the method of 
characteristics~\cite{deb97} (the mathematical details of this method 
have been outlined for the model problem presented in the Appendix) to 
get a solution that can be written as
\begin{equation}
\label{emin}
\frac{v^2}{2} - \frac{GM}{r} + \frac{\lambda^2}{2r^2} = {\mathcal F}
\left[ \frac{1}{r_0 + r\left(1 + v/a \right)} \exp \left( \frac{vr}{ar_0}
- \frac{at}{r_0} \right) \right]
\end{equation}
in which $r_0 = GM/a^2$, and $a$ itself is an integration constant 
deriving from the spatial part of the characteristic equations, and,
therefore, should, in general, be a function of time. The form
of the function $\mathcal{F}$ is to be determined from the physically
realistic initial condition $v=0$ at $t=0$ for all $r$, which will
render $\mathcal{F}$ as
\begin{equation}
\label{eff}
{\mathcal F}(\xi)=-\frac{GM\xi}{1-\xi r_0}+\frac{\lambda^2{\xi}^2}
{2\left(1-\xi r_0\right)^2} . 
\end{equation}
On examining the argument of $\mathcal{F}$ in (\ref{emin}) and then
studying its long-time behaviour, it will be seen from (\ref{eff}) that 
for $t \longrightarrow \infty$, the selected solution will correspond 
to the condition 
\begin{equation}
\label{select}
\frac{v^2}{2} - \frac{GM}{r} + \frac{\lambda^2}{2r^2} = 0 . 
\end{equation}

It can now be seen that prior to the evolution, the system had no bulk
motion and that the radial drift velocity was given flatly everywhere
by $v=0$. This, of course, gives the condition that initially the total
specific mechanical energy of the system was zero. Then at $t=0$ both
a gravitational mechanism is activated in this system and some angular
momentum is imparted to it. This will induce a potential $-GM/r$
everywhere, and at the same time start a rotational
motion, respectively. The system will then
start evolving in time, with the velocity $v$ at each point in space
evolving temporally according to (\ref{emin}). Finally the system will
restore itself to a steady state in such a manner that the total
specific mechanical energy at the end of the evolution
(for $t \longrightarrow \infty$)
will remain the same as at the beginning (at $t=0$), a condition that is
given by (\ref{select}), whose left hand side gives the sum of the
specific kinetic energy, the specific gravitational potential energy
and the specific rotational energy, respectively.
This sum is zero, and therefore, under the given initial condition,
this must be the steady state corresponding to the minimum possible total
specific energy of the system. Hence, this is the configuration that
is dynamically and non-perturbatively selected. It is now conceivable
that if the pressure term were to be taken into account, then the
solution that would be dynamically selected, would be the one that
would pass through the saddle point, since, as Bondi had analogously
conjectured for spherically symmetric accretion~\cite{bon52}, this
would be the one to satisfy the criterion of minimum energy.

Interestingly enough, the issue of the primacy of transonicity can also 
be addressed from a very different, and somewhat unlikely, perspective. 
It has been mentioned earlier that carrying out a linearised perturbative 
analysis in real time on stationary flows will not indicate any particular
solution to be favoured over all the others. This line of thinking may
now be subjected to a closer scrutiny. 

Under the condition of hydrostatic equilibrium in the vertical direction,
the time-dependent generalisation of the governing equations for an
axisymmetric pseudo-Schwarzschild disc is given by (\ref{dyneuler}),
as well as 
\begin{equation}
\label{surden}
\frac{\prt \Sigma}{\prt t}+ \frac{1}{r} \frac{\prt}{\prt r}
\left( \Sigma vr \right)=0
\end{equation}
in which the surface density of the thin disc $\Sigma$, is to be
expressed as $\Sigma \cong \rho H$~\cite{fkr02}.
Making use of (\ref{aitchpol})
and the polytropic relation $P = K \rho^{\gamma}$, (\ref{surden})
can be rendered as
\begin{equation}
\label{volden}
\frac{\prt}{\prt t} \left[\rho^{(\gamma +1)/2}\right]
+ \frac{\sqrt{\phi^{\prime}}}{r^{3/2}}
\frac{\prt}{\prt r} \left[ \rho^{(\gamma +1)/2} v
\frac{r^{3/2}}{\sqrt{\phi^{\prime}}} \right] =0 . 
\end{equation}

Defining a new variable
$f=\rho^{(\gamma +1)/2} v r^{3/2}/\sqrt{\phi^{\prime}}$, it is quite
obvious from the form of (\ref{volden}) that the stationary value
of $f$ will be a constant, $f_0$, which can be closely identified with
the matter flux rate. This follows a similar approach to spherically
symmetric flows established in earlier works~\cite{pso80,td92}. The 
present treatment, of course, pertains to a disc flow being driven by 
a general pseudo-Newtonian potential, $\phi (r)$. In this system, a 
perturbation prescription of the form $v(r,t) = v_0(r) + v^{\prime}(r,t)$ 
and $\rho (r,t) = \rho_0 (r) + \rho^{\prime}(r,t)$, will give, on
linearising in the primed quantities,
\begin{equation}
\label{effprime}
\frac{f^{\prime}}{f_0} = \left[\left( \frac{\gamma +1}{2} \right)
\frac{\rho^{\prime}}
{\rho_0} + \frac{v^{\prime}}{v_0} \right]
\end{equation}
with the subscript $0$ denoting stationary values in all cases. From
(\ref{volden}), it then becomes possible to set down the density
fluctuations $\rho^{\prime}$, in terms of $f^{\prime}$ as
\begin{equation}
\label{flucden}
\frac{\prt \rho^{\prime}}{\prt t} + \beta^2
\frac{v_0 \rho_0}{f_0} \left(\frac{\prt f^{\prime}}{\prt r}\right)=0
\end{equation}
with $\beta^2 = 2(\gamma +1)^{-1}$, as before.
Combining (\ref{effprime}) and (\ref{flucden}) will then render
the velocity fluctuations as
\begin{equation}
\label{flucvel}
\frac{\prt v^{\prime}}{\prt t}= \frac{v_0}{f_0}
\left(\frac{\prt f^{\prime}}{\prt t}+{v_0}
\frac{\prt f^{\prime}}{\prt r}\right)
\end{equation}
which, upon a further partial differentiation with respect to time,
will give
\begin{equation}
\label{flucvelder2}
\frac{{\prt}^2 v^{\prime}}{\prt t^2}=\frac{\prt}{\prt t} \left[
\frac{v_0}{f_0} \left(\frac{\prt f^{\prime}}{\prt t}\right) \right]
+ \frac{\prt}{\prt t} \left[ \frac{v_0^2}{f_0} \left(
\frac{\prt f^{\prime}}{\prt r}\right) \right] . 
\end{equation}

From (\ref{dyneuler}) the linearised fluctuating part could be
extracted as
\begin{equation}
\label{fluceuler}
\frac{\prt v^{\prime}}{\prt t}+ \frac{\prt}{\prt r}
\left( v_0 v^{\prime} + c_{\mathrm{s0}}^2
\frac{\rho^{\prime}}{\rho_0}\right) =0
\end{equation}
with $c_{\mathrm{s0}}$ being the speed of sound in the steady state.
Differentiating (\ref{fluceuler}) partially with respect to $t$,
and making use of (\ref{flucden}), (\ref{flucvel}) and
(\ref{flucvelder2}) to substitute for all the first and second-order
derivatives of $v^{\prime}$ and $\rho^{\prime}$, will deliver the result
\begin{eqnarray}
\label{interm}
\frac{\prt}{\prt t} \left[\frac{v_0}{f_0}
\left( \frac{\prt f^{\prime}}{\prt t}\right)\right]
+ \frac{\prt}{\prt t} \left[\frac{v_0^2}{f_0}
\left( \frac{\prt f^{\prime}}{\prt r}\right)\right]
&+& \frac{\prt}{\prt r} \left[\frac{v_0^2}{f_0}
\left( \frac{\prt f^{\prime}}{\prt t}\right)\right] \nonumber \\
&+& \frac{\prt}{\prt r} \left[\frac{v_0}{f_0}
\left(v_0^2 - \beta^2 c_{\mathrm{s0}}^2 \right)
\frac{\prt f^{\prime}}{\prt r}\right] = 0
\end{eqnarray}
all of whose terms can be ultimately rendered into a compact
formulation that looks like
\begin{equation}
\label{compact}
\prt_\mu \left( {\mathrm{f}}^{\mu \nu} \prt_\nu
f^{\prime}\right) = 0
\end{equation}
in which the Greek indices are made to run from $0$ to $1$, with
the identification that $0$ stands for $t$, and $1$ stands for $r$.
An inspection of the terms on the left hand side of (\ref{interm})
will then allow for constructing the symmetric matrix
\begin{equation}
\label{matrix}
{\mathrm{f}}^{\mu \nu } = \frac{v_0}{f_0}
\pmatrix
{1 & v_0 \cr
v_0 & v_0^2 - \beta^2 c_{\mathrm{s0}}^2} . 
\end{equation}

Now in Lorentzian geometry the d'Alembertian for a scalar in curved 
space is given in terms
of the metric ${\mathrm{g}}_{\mu \nu}$ by~\cite{vis98,blv05}
\begin{equation}
\label{alem}
\Delta \psi \equiv \frac{1}{\sqrt{-\mathrm{g}}}
\prt_\mu \left({\sqrt{-\mathrm{g}}}\, {\mathrm{g}}^{\mu \nu} \prt_\nu
\psi \right)
\end{equation}
with $\mathrm{g}^{\mu \nu}$ being the inverse of the matrix implied
by ${\mathrm{g}}_{\mu \nu}$. Comparing (\ref{compact}) and (\ref{alem})
it would be tempting to look for an exact equivalence between 
${\mathrm{f}}^{\mu \nu }$ and $\sqrt{-\mathrm{g}}\, {\mathrm{g}}^{\mu \nu}$.
This, however, cannot be done in a general sense. What can be 
appreciated, nevertheless, is that (\ref{compact}) gives an equation 
for $f^{\prime}$ which is of the type given by (\ref{alem}). The metrical
part of (\ref{compact}), as given by (\ref{matrix}), may then be 
extracted, and its inverse will incorporate the notion of a sonic 
horizon of an acoustic black hole when $v_0^2 = \beta^2 c_{\mathrm{s0}}^2$.
This point of view does not make for a perfect acoustic analogue model, 
but it has some similar features to the metric of a wave equation for a 
scalar field in curved space-time, obtained through a 
somewhat different approach, in which the velocity of an an irrotational, 
inviscid and barotropic fluid flow is first represented as the gradient 
of a scalar function $\psi$, i.e. ${\bf v}= -{\bf{\nabla}}\psi$, and 
then a perturbation is imposed on this scalar function~\cite{vis98,blv05}. 

The discussion above indicates that the physics of supersonic acoustic 
flows closely corresponds to many features of black hole physics. This 
closeness of form is very intriguing, as well as instructive. For a
black hole, infalling matter crosses the event horizon maximally, i.e.
at the greatest possible speed. By analogy the same thing may be said
of matter crossing the sonic horizon of a fluid flow. Indeed, this has
been a long-standing conjecture for the case of spherically symmetric 
accretion on to a point sink~\cite{bon52,gar79}. That this fact can be 
appreciated for the accretion problem through a perturbative
result, as given by (\ref{interm}), is quite remarkable. This is
because conventional wisdom would have it that one would be quite unable
to have any understanding of the special status of any inflow solution
solely through a perturbative technique~\cite{gar79}. It is the
transonic solution that crosses the sonic horizon at the greatest
possible rate, and the near similarity of form between (\ref{interm})
and (\ref{alem}) may very well be indicative of the primacy of the
transonic solution. If such an insight were truly to be had with the
help of the perturbation equation, then the perturbative linear stability
analysis might not have been carried out in vain after all.

\section{Concluding remarks}
\label{sec5}

The difficulties against transonicity, arising from the choice of purely
stationary equations, have been addressed through non-perturbative dynamics. 
This has been a relatively simple exercise to carry out in the Newtonian
construct of space and time. It may be readily appreciated, however, that 
a rigorously general relativistic flow will not lend itself so easily to
such a treatment as has been executed here for a pseudo-Newtonian system. 
Nevertheless, it should be helpful to note that particular 
pseudo-Newtonian systems~\cite{pw80} bear a very close resemblance to an 
actual general relativistic system~\cite{das02}. This closeness is a 
reason to believe that the non-perturbative approach to the selection 
of transonicity demonstrated in this paper may also be very relevant for 
general relativistic flows. 

As a final point, it may be mentioned that realistically speaking, a 
disc system should involve viscosity as a means of transporting angular
momentum, to make infall a sustained global process. However, 
in that event, viscosity will also bring about 
dissipation of energy in the flow, something that will make the minimum
energy criterion an ineffective instrument for identifying the transonic
solution. Besides this, viscosity will violate Lorentzian invariance,
which is very crucial for constructing an analogue gravity model. With
the breakdown of the Lorentzian invariance, the quest for an acoustic 
analogue of a black hole becomes a most difficult one, especially when 
the fluid system under study is compressible in nature. For incompressible 
flows, on the other hand, some studies have considered the issue of an
analogue gravity model vis-a-vis viscosity in a semi-quantitative manner. 
In analysing the outflow of a very shallow layer of water on a flat
surface, it has been discussed that the formation of an abrupt hydraulic 
jump in the flow 
is entirely due to viscosity, and many features of the jump itself can 
be closely connected to an acoustic white hole~\cite{vol05,sbr05}.  
But more importantly the basic properties of surface gravity waves 
in a shallow layer of water remain 
unchanged in the vicinity of the jump, inspite of viscosity~\cite{su02}. 
Going by this analogy, it is conceivable that the properties of an 
acoustic wave in a compressible flow (such as the accretion process is)
will likewise remain unaffected. 
This argument, however, will not make it directly possible to establish 
precise relations like (\ref{compact}) and (\ref{matrix}), which give 
a clear mathematical result 
to indicate that transonicity might be favoured. Yet transonicity will 
very likely be achieved because regardless of energy dissipation due 
to viscosity, the stationary matter flux rate will be conserved and 
the flow will continue to proceed at the greatest possible rate. 

\ack
This research has made use of NASA's Astrophysics Data System. The 
authors acknowledge some useful discussions with Rajaram Nityananda 
and Paul J. Wiita. 

\appendix
\section*{Appendix}
\setcounter{section}{1}

The model system that has been introduced here serves to show that
non-realisable separatrices passing through the saddle point in the 
stationary regime, can behave like proper physical flows, when the 
dynamics, as opposed to the statics, is to be followed. The dynamics 
of the field $y(x,t)$ is described as
\begin{equation}
\label{mod}
\frac{\partial y}{\partial t}+\left(y-x\right)\frac{\partial y}{\partial x}
=y+2x+x^2
\end{equation}
whose static limit leads to
\begin{equation}
\label{statmod}
\frac{\mathrm{d}y}{\mathrm{d}x} = \frac{y+2x+x^2}{y-x}
\end{equation}
and which, viewed as a dynamical system, is seen as
\begin{eqnarray}
\label{dynmod}
\frac{\mathrm{d}y}{\mathrm{d}\tau} &= y+2x+x^2 \nonumber \\
{\frac{\mathrm{d}x}{\mathrm{d}\tau}} &= y-x . 
\end{eqnarray}
In the $y$ --- $x$ space, the fixed points
$(x_{\mathrm c},y_{\mathrm c})$ are to be found at
$(0,0)$ and $(-3,-3)$. A linear stability
analysis of the fixed points in $\tau$ space gives the eigenvalues
$\Omega$, by $\Omega^2 = 1+2(x_{\mathrm c} +1)$. It is then 
easy to see that $(0,0)$ is a saddle point while $(-3,-3)$ is a 
centre-type point. 
This mathematical model is very much similar to the stationary
inviscid accretion disc driven by the Newtonian
potential, with two critical points, of which the outer one is 
a saddle point, while the inner one is centre-type. The integral 
curves of the model system are obtained from (\ref{statmod}) as 
\begin{equation}
\label{integmod}
y^2-2xy-2x^2- \frac{2}{3} x^3=c
\end{equation}
with the solutions passing through the saddle point being given by
$c=0$. 

To explore the temporal dynamics, and to obtain a solution to
(\ref{mod}), it would be necessary to apply the method of
characteristics~\cite{deb97}. This involves writing
\begin{equation}
\label{charac}
{\frac{\mathrm{d}t}{1}}={\frac{\mathrm{d}x}{y-x}}=
\frac{\mathrm{d}y}{y+2x+x^2} . 
\end{equation}
The subsequent task is to find two constants $c_1$ and $c_2$ from 
the above set and
the general solution of (\ref{mod}) would then be given by 
${c_1} = F(c_2)$, where the function $F$ is to be determined from 
the initial
conditions. It is easy to see that one of the constants of integration
is clearly the $c$ of (\ref{integmod}). Hence, writing ${c_1}=c$, 
and using (\ref{integmod}) in the first part of (\ref{charac}),
will give
\begin{equation}
\label{charsol}
\int \, \mathrm{d}t= \pm \int \frac{\mathrm{d}x}
{\sqrt{3x^2+\left(2/3\right)x^3+c}}
\end{equation}
which solves the problem in principle. To put this in a usable form,
the integration in (\ref{charsol}) would have to be carried out. 
For small $x$ (the most important region, since it
is near the saddle), the $x^3$ term may be left out to a good approximation.
Further, only the positive sign in the right hand side of (\ref{charsol})
is to be chosen by the physical argument that the system is to evolve
through a positive range of $t$ (time) values. Integration of
(\ref{charsol}) will then lead to the result
\begin{equation}
\label{timeinteg}
\left( x+ \sqrt{x^2+c/3} \right) e^{- \sqrt{3} t}={c_2}
\end{equation}
which will then make the solution of (\ref{mod}) look like
\begin{equation}
\label{gensol}
y^2-2xy-2x^2- \frac{2}{3} x^3=F\left (\left [x+
\sqrt{ \frac{\left(y-x \right)^2}{3}
- \frac{2}{9}x^3}\right ]e^{-{\sqrt 3}t} \right ) . 
\end{equation}

The evolution of the system is to be followed from the initial
condition that $y=0$ for all $x$ at $t=0$. Dropping the $x^3$ term again
for small $x$, will allow for determining the form of the function
$F$ as $F(z)=-3(2- \sqrt{3})z^2$. The solution, consequently, becomes
\begin{equation}
\label{solinit}
y^2-2xy-2x^2- \frac{2}{3} x^3=-3\left(2- \sqrt{3}\right)
\left [ x+ \sqrt{ \frac{\left(y-x \right)^2}{3}-
\frac{2}{9}x^3}\right ]^2 e^{-2{\sqrt{3}}t}
\end{equation}
and as $t{\longrightarrow}{\infty}$, the steady solution that
would be selected would be
\begin{equation}
\label{integmod2}
y^2-2xy-2x^2- \frac{2}{3} x^3=0
\end{equation}
which is actually the equation for the separatrices.
It is worth stressing the remarkable feature of this result. The
evolution started under conditions far removed from transonicity. In
fact, it started as $y=0$ for all $x$ (in the vicinity of the origin of
coordinates) at $t=0$. The evolution proceeded through a myriad of
possible steady state solutions (all arguably stable under a linear
stability analysis) and then in the stationary limit, selected the
separatrices. This is a convincing demonstration that it is in principle
possible for apparently non-realisable separatrices in the steady regime,
to become eminently realisable physically, when the temporal evolution of
the system is followed.


\section*{References}

\begin{thebibliography}{99}

\bibitem{skc90} Chakrabarti S K 1990 {\it Theory of Transonic 
Astrophysical Flows} (Singapore: World Scientific)

\bibitem{c89} Chakrabarti S K 1989 {\it ApJ} {\bf 347} 365

\bibitem{bon52} Bondi H 1952 {\it MNRAS} {\bf 112} 195

\bibitem{rb02} Ray A K and Bhattacharjee J K 2002 {\it Phys. Rev. E} 
{\bf 66} 066303

\bibitem{gar79} Garlick A R 1979 {\it A\&A} {\bf 73} 171

\bibitem{az81} Abramowicz M A and Zurek W H 1981 {\it ApJ} {\bf 246} 314

\bibitem{fuk87} Fukue J 1987 {\it PASJ} {\bf 39} 309

\bibitem{ky94} Kafatos M and Yang R X 1994 {\it MNRAS} {\bf 268} 925

\bibitem{yk95} Yang R X and Kafatos M 1995 {\it A\&A} {\bf 295} 238

\bibitem{par96} Pariev V I 1996 {\it MNRAS} {\bf 283} 1264

\bibitem{la97} Lasota J-P and Abramowicz M A 1997 {\it Classical and
Quantum Gravity} {\bf 14} A237

\bibitem{lyyy97} Lu J F, Yu K N, Yuan F and Young E C M 1997 
{\it A\&A} {\bf 321} 665

\bibitem{pa97} Peitz J and Appl S 1997 {\it MNRAS} {\bf 286} 681

\bibitem{das02} Das T K 2002 {\it ApJ} {\bf 577} 880

\bibitem{bdw04} Barai P, Das T K and Wiita P J 2004 {\it ApJ} 
{\bf 613} L49

\bibitem{das04} Das T K 2004 {\it MNRAS} {\bf 349} 375

\bibitem{abd06} Abraham H, Bili\'c N and Das T K 2006 
{\it Classical and Quantum Gravity} {\bf 23} 2371

\bibitem{dbd06} Das T K, Bili\'c N and Dasgupta S 2006
{\tt arXiv:astro-ph}/0604477

\bibitem{js99} Jordan D W and Smith P 1999 {\it Nonlinear Ordinary
Differential Equations} (Oxford: Oxford University Press)

\bibitem{ap03} Afshordi N and Paczy\'nski B 2003 {\it ApJ} {\bf 592} 354

\bibitem{crd06} Chaudhury S, Ray A K and Das T K 2006 {\it MNRAS} 
{\bf 373} 146

\bibitem{ray03} Ray A K 2003 {\it MNRAS} {\bf 344} 1085

\bibitem{mkfo84} Matsumoto R, Kato S, Fukue J and Okazaki A T 1984 
{\it PASJ} {\bf 36} 71

\bibitem{fkr02} Frank J, King A and Raine D 2002 {\it Accretion Power in
Astrophysics} (Cambridge: Cambridge University Press) 

\bibitem{bdp93} Bohr T, Dimon P and Putkaradze V 1993 {\it Journal of 
Fluid Mechanics} {\bf 254} 635

\bibitem{lt80} Liang E P T and Thompson K A 1980 {\it ApJ} {\bf 240} 271

\bibitem{deb97} Debnath L 1997 {\it Nonlinear Partial Differential 
Equations for Scientists and Engineers} (Boston: Birkh{\"a}user)

\bibitem{pso80} Petterson J A, Silk J and Ostriker J P 1980
{\it MNRAS} {\bf 191} 571

\bibitem{td92} Theuns T and David M 1992 {\it ApJ} {\bf 384} 587

\bibitem{vis98} Visser M 1998 {\it Classical and Quantum Gravity}
{\bf 15} 1767

\bibitem{blv05} Barcel{\'o} C, Liberati S and Visser M 2005 
{\tt arXiv:gr-qc}/0505065

\bibitem{pw80} Paczy\'nski B and Wiita P J  1980 {\it A\&A} {\bf 88} 23

\bibitem{vol05} Volovik G E 2005 {\it JETP Letters} {\bf 82} 624

\bibitem{sbr05} Singha S B, Bhattacharjee J K and Ray A K 2005 
{\it European Physical Journal B} {\bf 48} 417

\bibitem{su02} Sch{\"u}tzhold R and Unruh W 2002 {\it Phys. Rev. D} 
{\bf 66} 044019

\end{thebibliography}
\end{document}